\documentclass[showpacs,10pt,twocolumn,prb]{revtex4-1}
\usepackage{amsmath}
\usepackage{amssymb}
\usepackage{graphics}
\usepackage{epsfig}

\begin{document}


\title{Antiferromagnetism in semiconducting KFe$_{0.85}$Ag$_{1.15}$Te$_{2}$
single crystals}
\author{Hechang Lei, Emil S. Bozin, Kefeng Wang, and C. Petrovic}
\affiliation{Condensed Matter Physics and Materials Science Department, Brookhaven
National Laboratory, Upton, NY 11973, USA}
\date{\today}

\begin{abstract}
We have synthesized single crystals of K$_{1.00(3)}$Fe$_{0.85(2)}$Ag$%
_{1.15(2)}$Te$_{2.0(1)}$. The materials crystallizes in the ThCr$_{2}$Si$_{2}
$ structure with I4/mmm symmetry and without K and Fe/Ag deficiencies,
unlike in K$_{x}$Fe$_{2-y}$Se$_{2}$ and K$_{x}$Fe$_{2-y}$S$_{2}$. Transport, magnetic, and heat capacity measurements indicate that KFe$_{0.85}
$Ag$_{1.15}$Te$_{2}$ is a semiconductor with long-range antiferromagnetic
transition at $T_{N}$ = 35 K.
\end{abstract}

\maketitle

The discovery of superconductivity in LaFeAsO$_{1-x}$F$_{x}$\cite{Kamihara}
has stimulated substantial interest in iron based high temperature
superconductors (Fe-HTS). Until now, several Fe-HTS were discovered. They
can be divided into two classes. The first class are iron pnictide materials.%
\cite{Kamihara}$^{-}$\cite{Wang XC} They contain two dimensional FePn (Pn =
pnictogens) tetrahedron layers and atomic sheets (e.g. Ba, K) or complex
blocks (e.g. La-O(F)) along the c axis. Another class are binary iron
chalcogenides FeCh (Ch = chalcogens, FeCh-11 type).\cite{Hsu FC}$^{-}$\cite%
{Mizuguchi} In contrast to the diversity of iron pnictide superconductors,
FeCh-11 type materials do not have any atomic or complex layers between
puckered FeCh sheets.

Very recently, the discovery of A$_{x}$Fe$_{2-y}$Se$_{2}$ (A = K, Rb, Cs and
Tl, FeCh-122 type) with $T_{c}\approx $ 30 K raised $T_{c}$ in Fe-HTS by
introducing alkali metal atomic layers between FeCh sheets.\cite{Guo}$^{-}$%
\cite{Fang MH} Further studies indicate that in the new superconductors the $%
T_{c}$ gets enhanced when compared to FeCh-11 materials, but there is also a
set of distinctive physical properties. FeCh-122 materials
are close to the metal-semiconducting crossover and antiferromagnetic (AFM)
order.\cite{Guo}$^{-}$\cite{Fang MH} This is in contrast to other
superconductors which are in close proximity to the spin density wave state.%
\cite{Cruz} Fermi surface in FeCh-122 type Fe-HTS contains only electron
like sheets without nesting features found in most other Fe-HTS.\cite{Zhang
Y}

On the other hand, superconductivity in FeCh-11 materials is quite robust
with respect to anion change as seen on the example of FeSe$_{1-x}$, FeTe$%
_{1-x}$Se$_{x}$ and FeTe$_{1-x}$S$_{x}$.\cite{Hsu FC}$^{-}$\cite{Mizuguchi}
However, in FeCh-122 compounds, superconductivity is only observed in A$_{x}$%
Fe$_{2-y}$Se$_{2}$ or A$_{x}$Fe$_{2-y}$Se$_{2-z}$S$_{z}$,\cite{Lei HC1}
while pure K$_{x}$Fe$_{2-y}$S$_{2}$ is a semiconductor with spin glass
transition at low temperature.\cite{Lei HC2} Moreover, the theoretical
calculation indicates that the hypothetical KFe$_{2}$Te$_{2}$, if
synthesized, would have higher $T_{c}$\ than K$_{x}$Fe$_{2-y}$Se$_{2}$.\cite%
{Shein} Therefore, synthesis and examination of physical properties of
FeCh-122 materials containing FeTe layers could be very instructive.

In this work, we report discovery of K$_{1.00(3)}$Fe$_{0.85(2)}$Ag$%
_{1.15(2)} $Te$_{2.0(1)}$ single crystals. The resistivity and magnetic
measurements indicate that this compound has the semiconducting long range
antiferromagnetic (AFM) order at low temperature with no superconductivity
down to 1.9 K.

Single crystals of K(Fe,Ag)$_{2}$Te$_{2}$ were grown by self-flux method
reported elsewhere in detail\cite{Lei HC2}$^{,}$\cite{Lei HC3} with nominal
composition K:Fe:Ag:Te = 1:1:1:2. Single crystals with typical size 5$\times
$5$\times $2 mm$^{3}$ can be grown. Powder X-ray diffraction (XRD) data were
collected at 300 K using 0.3184 \AA\ wavelength radiation (38.94 keV) at X7B
beamline of the National Synchrotron Light Source. The average stoichiometry
was determined by examination of multiple points using an energy-dispersive
x-ray spectroscopy (EDX) in a JEOL JSM-6500 scanning electron microscope.
Electrical transport measurements were performed using a four-probe
configuration on rectangular shaped polished single crystals with current
flowing in the ab-plane of tetragonal structure. Thin Pt wires were attached
to electrical contacts made of silver paste. Electrical transport, heat
capacity, and magnetization measurements were carried out in Quantum Design
PPMS-9 and MPMS-XL5.

Figure 1(a) shows powder XRD result and structural refinements of K(Fe,Ag)$%
_{2}$Te$_{2}$\ using General Structure Analysis System (GSAS).\cite{Larson}$%
^{,}$\cite{Toby} It can be seen that all reflections can be indexed in the
I4/mmm space group. The refined structure parameters are listed in Table 1.
The determined lattice parameters are a = 4.3707(9) \r{A}\ and c =
14.9540(8) \r{A}, which are reasonably smaller than those of CsFe$_{x}$Ag$%
_{2-x}$Te$_{2}$ (a = 4.5058(4) \r{A}\ and c = 15.4587(8) \r{A}),\cite{Li J}
but much larger than those of K$_{x}$Fe$_{2-y}$Se$_{2}$ and K$_{x}$Fe$_{2-y}$%
S$_{2}$,\cite{Guo}$^{,}$\cite{Lei HC2} due to smaller ionic size of K$^{+}$
than Cs$^{+}$ and larger size of Ag$^{+}$ and Te$^{2-}$ than Fe$^{2+}$ and Se%
$^{2-}$(S$^{2-}$). On the other hand, larger a-axis lattice parameter
indicates that the Fe plane is stretched in K(Fe,Ag)$_{2}$Te$_{2}$ when
compared to FeTe.\cite{Rodriguez} The crystal structure of K(Fe,Ag)$_{2}$Te$%
_{2}$ is shown in Fig. 1(b), where antifluorite-type Fe/Ag-Te layers and K
cation layers are stacked alternatively along the c axis. XRD pattern of a
single crystal (Fig. 1(c)) reveals that the crystal surface is normal to the
c axis with the plate-shaped surface parallel to the ab-plane. Fig. 1(d)
presents the EDX spectrum of a single crystal, which confirms the presence
of the K, Fe, Ag, and Te. The average atomic ratios determined from EDX are
K:Fe:Ag:Te = 1.00(3):0.85(2):1.15(2):2.0(1). The value of Fe/(Ag+Fe)
determined from XRD fitting (0.38) is close to that obtained from EDX
(0.43). It suggests that Te compound prefers to contain more Ag. This might
explain why pure KFe$_{2}$Te$_{2}$ can not form since large Ag$^{+}$ ions
have to be introduced in order to match the rather large Te$^{2-}$ anions
and keep the stability of the structure. On the other hand, it should be
noted that there are no K or Fe/Ag deficiencies in K(Fe,Ag)$_{2}$Te$_{2}$.
This is rather different from K$_{x}$Fe$_{2-y}$Se$_{2}$ and K$_{x}$Fe$_{2-y}$%
S$_{2}$.\cite{Guo}$^{,}$\cite{Lei HC2} Moreover, synchrotron powder X-ray
refinement and EDX were consistent with either stoichiometric Te or not more
than 5\% vacancies (i.e., Te$_{1.9}$)

\begin{figure}[tbp]
\centerline{\includegraphics[scale=0.4]{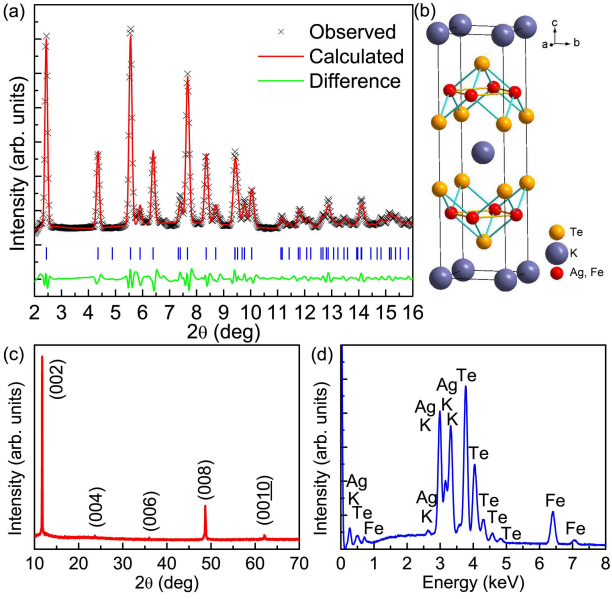}} \vspace*{-0.3cm}
\caption{(a) Powder XRD patterns of KFe$_{0.85}$Ag$_{1.15}$Te$%
_{2} $. (b) Crystal structure of KFe$_{0.85}$Ag$_{1.15}$Te$_{2}$%
. The big blue, small red and medium orange balls represent K, Fe/Ag, and Te
ions. (c) Single crystal XRD of KFe$_{0.85}$Ag$_{1.15}$Te$_{2}$.
(d) The EDX spectrum of a single crystal.}
\end{figure}

\begin{table}[tbp]\centering%
\caption{Structural parameters for K(Fe,Ag)$_{2}$ Te$_{2}$ at room
temperature. Values in brackets give the number of equivalent distances or
angles of each type. The occupancies of K and Te are fixed during fitting.}%
\begin{tabular}{cccccc}
\hline\hline
\multicolumn{3}{c}{Chemical Formula} & \multicolumn{3}{c}{K$_{1.00(3)}$Fe$%
_{0.85(2)}$Ag$_{1.15(2)}$Te$_{2.0(1)}$} \\
\multicolumn{3}{c}{Space Group} & \multicolumn{3}{c}{I4/mmm} \\
\multicolumn{3}{c}{a (\AA )} & \multicolumn{3}{c}{4.3707(9)} \\
\multicolumn{3}{c}{c (\AA )} & \multicolumn{3}{c}{14.9540(8)} \\
\multicolumn{3}{c}{V (\AA $^{3}$)} & \multicolumn{3}{c}{285.7(1)} \\ \hline
\multicolumn{3}{c}{Interatomic Distances (\AA )} & \multicolumn{3}{c}{Bond
Angles ($^{\circ }$)} \\
\multicolumn{2}{c}{d$_{Fe/Ag-Fe/Ag}$ [4]} & 3.0906(4) & \multicolumn{2}{c}{
Te-Fe/Ag-Te [2]} & 104.44(3) \\
\multicolumn{2}{c}{d$_{Fe/Ag-Te/Ag}$ [4]} & 2.7651(5) & \multicolumn{2}{c}{
Te-Fe/Ag-Te [4]} & 112.05(4) \\
\multicolumn{2}{c}{Anion Heights (\AA )} & 1.694(7) &  &  &  \\ \hline
Atom & x & y & z & Occ & U$_{iso}$ (\AA $^{2}$) \\
K & 0 & 0 & 0 & 1.00 & 0.056(5) \\
Fe & 0.5 & 0 & 0.25 & 0.76(8) & 0.035(7) \\
Ag & 0.5 & 0 & 0.25 & 1.24(8) & 0.035(7) \\
Te & 0.5 & 0.5 & 0.1367(5) & 1.00 & 0.034(4) \\ \hline\hline
\end{tabular}%
\label{TableKey}%
\end{table}%

\begin{figure}[tbp]
\centerline{\includegraphics[scale=0.4]{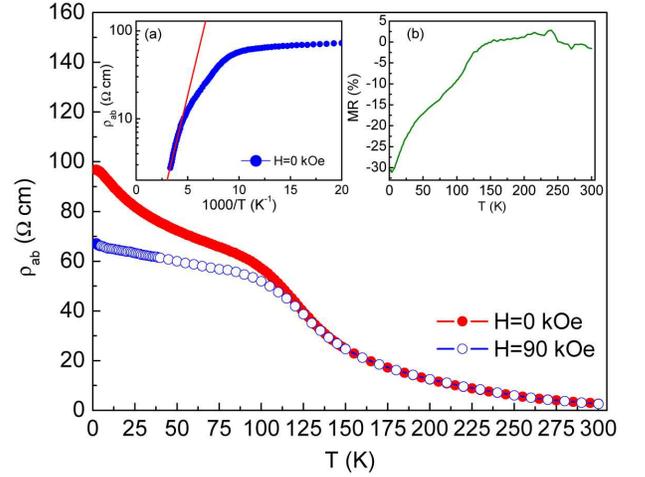}} \vspace*{-0.3cm}
\caption{Temperature dependence of the in-plane resistivity $\protect\rho %
_{ab}(T)$ of the KFe$_{0.85}$Ag$_{1.15}$Te$_{2}$ single crystal with $H$ = 0
(closed red circle) and 90 kOe (open blue square, H$\Vert $c). Inset (a)
shows the fitted result using thermal activation model for $\protect\rho %
_{ab}(T)$ at zero field where the red line is the fitting curve. Inset (b)
exhibits the temperature dependence of MR(T) for KFe$_{0.85}$Ag$_{1.15}$Te$%
_{2}$.}
\end{figure}

Figure 2 shows the temperature dependence of the in-plane resistivity $\rho
_{ab}$(T) of the KFe$_{0.85}$Ag$_{1.15}$Te$_{2}$ single crystal for $H$ = 0
and 90 kOe. The resistivity increases with decreasing the temperature with a
"shoulder" appearing around 100 K. The room-temperature value $\rho _{ab}$
is about 2.7 $\Omega \cdot $cm, which is much larger than in superconducting
K$_{x}$Fe$_{2-y}$Se$_{2}$ and semiconducting K$_{x}$Fe$_{2-y}$S$_{2}$.\cite%
{Lei HC2}$^{,}$\cite{Lei HC3} The semiconducting behavior might be related
to the random distribution of Fe and Ag ions in the Fe/Ag plane which
induces a random scattering potential, similar to the effect of Fe
deficiency in the FeSe or FeS plane.\cite{Lei HC2}$^{,}$\cite{Wang DM} By
fitting the $\rho _{ab}$(T) at high temperature using the thermal activation
model $\rho =\rho _{0}\exp (E_{a}/k_{B}T)$, where $\rho _{0}$ is a prefactor
and $k_{B}$ is Boltzmann's constant (inset (a) of Fig. 2), we obtained $\rho
_{0}$ = 71(6) m$\Omega \cdot $cm and the activation energy $E_{a}$ = 96(2)
meV in the temperature range above 200 K, which is larger than that of K$_{x}
$Fe$_{2-y}$S$_{2}$.\cite{Lei HC2} KFe$_{0.85}$Ag$_{1.15}$Te$_{2}$ exhibits
large magnetoresistance (MR $=[\rho (H)-\rho (0)]/\rho (0)$) below about 100
K where the shoulder appears. As shown in the inset (b) of Fig. 2, the
negative MR is about 30\% at 1.9 K for $H$ = 90 kOe. This behavior is
distinctively different from K$_{x}$Fe$_{2-y}$S$_{2}$, which does not show
any MR in measured temperature range.\cite{Lei HC2}

\begin{figure}[tbp]
\centerline{\includegraphics[scale=0.45]{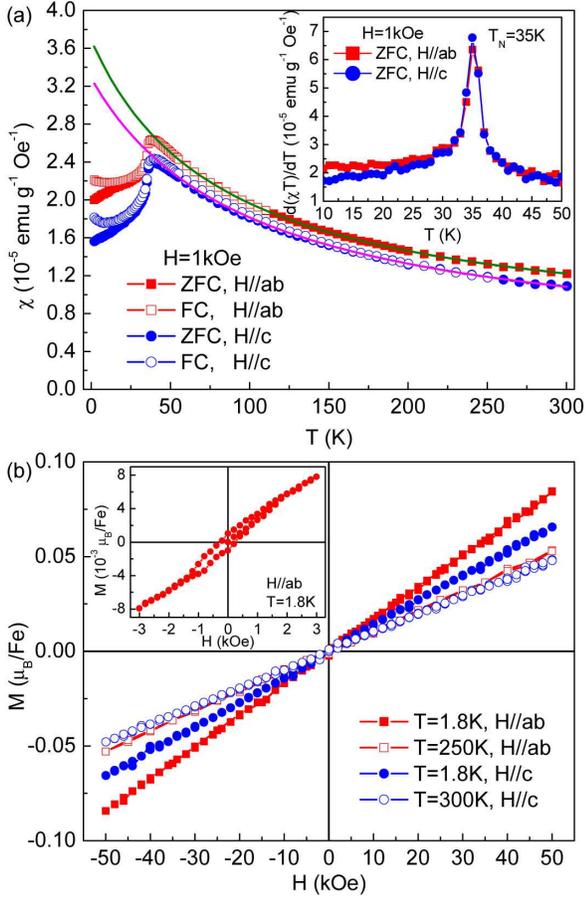}} \vspace*{-0.3cm}
\caption{(a) Temperature dependence of DC magnetic susceptibility $\protect%
\chi $(T) with the applied field $H$ = 1 kOe along ab plane and c axis below
300 K under ZFC and FC mode. The inset shows $d(\protect\chi T)/dT$ result
for both field directions. (b) Isothermal magnetization hysteresis loops
M(H) for H$\Vert $ab and H$\Vert $c at various temperatures.}
\end{figure}

Figure 3(a) presents the temperature dependence of magnetic susceptibility $%
\chi (T)$ of the KFe$_{0.85}$Ag$_{1.15}$Te$_{2}$ single crystal for $H$ = 1
kOe along the ab plane and the c axis below 300 K with zero-field-cooling
(ZFC) and field-cooling (FC). The $\chi _{ab}(T)$ is slightly larger than $%
\chi _{c}(T)$ and above 50 K, both can be fitted very well using Curie-Weiss
law $\chi (T)=\chi _{0}+C/(T-\theta )$, where $\chi _{0}$ includes core
diamagnetism, van Vleck and Pauli paramagnetism, $C$ is Curie constant and $%
\theta $ is the Curie-Weiss temperature (solid lines in Fig. 3(a)). The
fitted parameters are $\chi _{0}$ = 5.46(7)$\times $10$^{-6}$ emu g$^{-1}$ Oe%
$^{-1}$, $C$ = 2.58(3)$\times $10$^{-3}$ emu g$^{-1}$ Oe$^{-1}$ K, and $%
\theta $ = -82(1) K for H$\Vert $ab and $\chi _{0}$ = 3.5(1)$\times $10$%
^{-6} $ emu g$^{-1}$ Oe$^{-1}$, $C$ = 2.92(5)$\times $10$^{-3}$ emu g$^{-1}$
Oe$^{-1}$ K, and $\theta $ = -100(2) K for\ H$\Vert $c. The above values of $%
C$ correspond to an effective moment of $\mu _{eff}$ = 3.60(2) $\mu _{B}/Fe$
and 3.83(3) $\mu _{B}/Fe$ for H$\Vert $ab and H$\Vert $c, respectively. The
values of $\mu _{eff}$ are smaller than for free Fe$^{2+}$ ions (4.7 $\mu
_{B}/Fe$) and Fe$_{1+x}$Te (4.9 $\mu _{B}/Fe$),\cite{Liu} but slightly
larger than in K$_{x}$Fe$_{2-y}$Se$_{2}$ (3.31 $\mu _{B}/Fe$).\cite{Bao W}

We observe sharp drops below 35 K in both ZFC and FC curves, associated with
the onset of long-range AFM order. The $T_{N}$ = 35 K is determined from the
peak of $d(\chi T)/dT$ (inset of Fig. 3(a)).\cite{Fisher} It should be noted
that antiferromagnetism below 35 K and Curie-Weiss paramagnetism at higher
temperature are obviously different from K$_{x}$Fe$_{2-y}$Se$_{2}$ and K$%
_{x} $Fe$_{2-y}$S$_{2}$.\cite{Lei HC1} Fig. 3 (b) shows the magnetization
loops for both field directions at various temperatures. It can be seen that
all M-H loops exhibit almost linear field dependence and M(H) curve exhibits
a very small hysteresis at 1.8 K with the small coercive field ($\sim $ 260
Oe).

\begin{figure}[tbp]
\centerline{\includegraphics[scale=0.9]{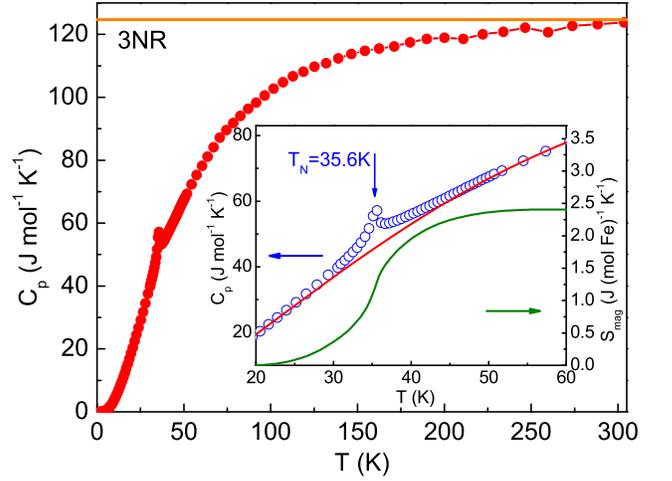}} \vspace*{-0.3cm}
\caption{Temperature dependence of heat capacity for KFe$_{0.85}$Ag$_{1.15}$%
Te$_{2}$ single crystal. The orange solid line represents the classical
value according to Dulong-Petit law at high temperature. The inset shows the
enlarged area near magnetic transition region of $C_{p}-T$. The red solid
curve represents the lattice contribution, fitted by a polynomial. The right
label denotes the magnetic entropy associated with the AFM transition.}
\end{figure}

Figure 4 shows the temperature dependence of heat capacity $C_{p}$ for KFe$%
_{0.85}$Ag$_{1.15}$Te$_{2}$ single crystals measured between T = 1.95 and
300 K in a zero magnetic field. At high temperature heat capacity approaches
the Dulong-Petit value of 3NR, where N is the atomic number in the chemical
formula (N = 5) and R is the gas constant (R = 8.314 J mol$^{-1}$ K$^{-1}$).
On the other hand, at the low temperature $C_{p}(T)$ curve can be fitted
solely by a cubic term $\beta T^{3}$. By neglecting antiferromagnon
contribution,\cite{Kranendonk} from the fitted value of $\beta $ = 3.11(2)
mJ mol$^{-1}$ K$^{-4}$, the Debye temperature is estimated to be $\Theta
_{D} $ = 146.2(3) K using the formula $\Theta _{D}$ = $(12\pi ^{4}NR/5\beta
)^{1/3}$. This is much smaller than $\Theta _{D}$ of K$_{x}$Fe$_{2-y}$Se$%
_{2} $ and K$_{x}$Fe$_{2-y}$S$_{2}$ at least partially because of larger
atomic mass of Ag and Te in KFe$_{0.85}$Ag$_{1.15}$Te$_{2}$.\cite{Lei HC2}$%
^{,}$\cite{Zeng B}

A $\lambda $-type anomaly at $T_{N}$ = 35.6 K (shown in the inset of Fig. 4)
confirms the bulk nature of the AFM order observed in the magnetization
measurement shown in Fig. 3. The transition temperature is consistent with
the values determined from $d(\chi T)/dT$ (35 K). Assuming that the total
heat capacity consists of phonon ($C_{ph}$) and magnetic ($C_{mag}$)
components, the $C_{mag}$ can be estimated by the subtraction of $C_{ph}$.
Consequently, the magnetic entropy ($S_{mag}$) can be calculated using the
integral $S_{mag}(T)=\int\limits_{0}^{T}C_{mag}/TdT$. Because of the failure
of Debye model at T $>$ $\Theta _{D}$, we estimated the lattice specific
heat by fitting a polynomial to the $C_{p}(T)$ curve at temperatures well
away from $T_{N}$. The obtained $S_{mag}$ is about 2.4 J (mol Fe)$^{-1}$ K$%
^{-1}$ up to 60 K, which is only 18\% of theoretical value (Rln5 = 13.4 J
(mol Fe)$^{-1}$ K$^{-1}$ for high spin state Fe$^{2+}$ ions). Note that only
about 1 J (mol Fe)$^{-1}$ K$^{-1}$ is released below $T_{N}$. This
discrepancy may originate from an incorrect estimation of the lattice
contribution to $C_{ph}(T)$ which can lead to reduced $S_{mag}(T)$ or a
probable short-range order that may exist above the bulk three-dimensional
AFM order occurring at $T_{N}$. This could also be supported by a much
smaller $T_{N}$ = 35 K compared to the Curie-Weiss temperature $\theta $ =
-100(2) K for H$\parallel $c.

There are two origins could induce the negative MR effect in semiconductors: the
reduction in spin disorder scattering due to the alignment of moments under
a field, and the reduction of the gap arising from the splitting of the up-
and down-spin subbands. The existence of AFM interaction could be related to
this negative MR\ effect. The temperature where MR effect becomes obvious
is consistent with the Curie-Weiss temperature $\theta $, which imply this
MR effect could be related to the AFM interaction and due to the reduction
in spin disorder scattering with field. When compared to K$_{x}$Fe$_{2-y}$Se$%
_{2}$, substitution of Ag has an important influence on the magnetic and
transport properties. It could reduce the exchange interaction between Fe
atoms and thus suppress the $T_{N}$ of KFe$_{0.85}$Ag$_{1.15}$Te$_{2}$
significantly. On the other hand, because of the near absence of vacancies
in KFeCuS$_{2}$ and similar valence between Cu and Ag,\cite{Oledzka} it is
more meaningful to compare the physical properties between KFeCuS$_{2}$ and
KFe$_{0.85}$Ag$_{1.15}$Te$_{2}$. The former has the larger $E_{a}$ and
room-temperature resistivity than the latter. This could be due to the
smaller ionic sizes of Cu and S when compared to Ag and Te, which might lead
to the smaller orbital overlap increasing $E_{a}$ and resistivity. Both
compounds exhibit Curie-Weiss law above 50 K and the fitted Curie-Weiss
temperature are also similar (176 K for KFeCuS$_{2}$). Moreover, KFeCuS$_{2}$
shows magnetic transition at 40 K, which is very close to the $T_{N}$ of KFe$%
_{0.85}$Ag$_{1.15}$Te$_{2}$. However, the transition of KFeCuS$_{2}$ is spin
glass-like, in contrast to the long-range AFM order of KFe$_{0.85}$Ag$_{1.15}
$Te$_{2}$. This implies that the distribution of Ag in KFe$_{0.85}$Ag$_{1.15}
$Te$_{2}$ may be different from Cu in KFeCuS$_{2}$ which results in
different magnetic ground state configuration with similar interaction
strength.

In summary, we successfully synthesized the K$_{1.00(3)}$Fe$_{0.85(2)}$Ag$%
_{1.15(2)}$Te$_{2.0(1)}$ single crystals with ThCr$_{2}$Si$_{2}$ structure,
identical to K$_{0.8}$Fe$_{2-y}$Se$_{2}$ at 600 K. Crytal structure and
composition analysis indicate that there are no K, Fe/Ag and Te vacancies
within 3, 2, or 5 atomic \%, respectively. Transport, magnetic, and thermal
measurements indicate that the KFe$_{0.85}$Ag$_{1.15}$Te$_{2}$ is a
semiconductor with long-rang AFM order below 35 K.

We thank John Warren for help with scanning electron microscopy measurements
and Jonathan Hanson for help with X-ray measurements. Work at Brookhaven is
supported by the U.S. DOE under Contract No. DE-AC02-98CH10886 and in part
by the Center for Emergent Superconductivity, an Energy Frontier Research
Center funded by the U.S. DOE, Office for Basic Energy Science

\end{document}